\documentclass[galaxies,article,accept,moreauthors,pdftex,10pt,a4paper]{Definitions/mdpi}

\firstpage{1}
\pubvolume{xx}
\issuenum{1}
\articlenumber{5}
\pubyear{2019}
\copyrightyear{2019}
\history{Received: 11 January 2019; Accepted: 14 February 2019; Published: date}

\newcommand{\td}[1]{\, \mbox{d} #1 \,}

\newcommand{\fermi}{\textit{Fermi}-LAT}
\newcommand{\g}{\ensuremath{\gamma}}
\newcommand{\cta}{CTA\,102} 
\newcommand{\zred}{z_{\rm red}}

\newcommand{\logb}[1]{\ln{\left( #1 \right)}}
\newcommand{\p}{^{\prime}}

\newcommand{\E}[1]{\times 10^{#1}}

\Title{The Long-Lasting Activity in the Flat Spectrum Radio Quasar (FSRQ) CTA~102}

\Author{Michael Zacharias $^{1,2,}$*\orcidA{}, Markus B\"ottcher $^{2}$\orcidB{}, Felix Jankowsky $^3$, Jean-Philippe Lenain $^4$\orcidC, Stefan J. Wagner $^3$ and Alicja Wierzcholska $^5$\orcidD{}
}

\AuthorNames{Michael Zacharias, Markus B\"ottcher, Felix Jankowsky, Jean-Philippe Lenain, Stefan J. Wagner and Alicja Wierzcholska
}

\address{%
$^{1}$ \quad Ruhr Astroparticle and Plasma Physics Center (RAPP Center), Insitut f\"ur theoretische Physik IV, Ruhr-Universit\"at Bochum, D-44780 Bochum, Germany  \\
$^{2}$ \quad Centre for Space Research, North-West University, Potchefstroom 2520, South Africa; markus.bottcher@nwu.ac.za   \\
$^3$ \quad Landessternwarte, Universit\"at Heidelberg, K\"onigstuhl, D-69117 Heidelberg, Germany; f.jankowsky@lsw.uni-heidelberg.de (F.J.); s.wagner@lsw.uni-heidelberg.de (S.J.W.) \\
$^4$ \quad Sorbonne Universit\'e, Universit\'e Paris Diderot, Sorbonne Paris Cit\'e, CNRS/IN2P3, Laboratoire de Physique Nucl\'eaire et de Hautes Energies, LPNHE, 4 Place Jussieu, F-75252 Paris, France; jean-philippe.lenain@lpnhe.in2p3.fr \\
$^5$ \quad Institute of Nuclear Physics, Polish Academy of Sciences, PL-31342 Krakow, Poland; alicja.wierzcholska@ifj.edu.pl
}

\corres{Correspondence: mz@tp4.rub.de or mzacharias.phys@gmail.com
}

\abstract{The flat spectrum radio quasar CTA 102 ($z = 1.032$) {went through a tremendous phase of variability}. Since early 2016 the gamma-ray flux level has been significantly higher than in previous years. It was topped by a four month long giant outburst, where peak fluxes were more than 100 times higher than the quiescence level. Similar trends are observable in optical and X-ray energies. We have explained the giant outburst as the ablation of a gas cloud by the relativistic jet that injects additional matter into the jet and can self-consistently explain the long-term light curve. Here, we argue that the cloud responsible for the giant outburst is part of a larger system that collides with the jet and is responsible for the years-long activity in CTA 102. 
} 


\begin{document}


\section{Introduction}
Detailed monitoring across the electromagnetic spectrum, such as the programs by, e.g., \fermi\ in the high energy (HE) $\gamma$-ray domain  or ATOM at optical frequencies (see, e.g., the reports by D. Thompson \cite{t18}, and J.-P Lenain \cite{l18} in this issue), is an essential tool to find new and unexpected behavior in blazars. Blazars, the relativistically beamed version of active galactic nuclei~\cite{br74}, are highly variable objects which can double their fluxes in all energy bands from radio to $\gamma$-rays and at all time scales from years down to minutes. Flux variations up to factors 100 are also observed, which usually happen on shorter time scales. And while this seems to be a common rule that bright flares tend to be short, there can be exceptions.

One of these exception is the blazar \cta, which is a flat spectrum radio quasar (FSRQ) at a redshift of $\zred=1.032$. As an FSRQ, it contains strong, broad emission lines pointing to the existence of a broad-line region (BLR). The BLR luminosity is $L_{\rm BLR}\p = 4.14\times 10^{45}\,$erg/s and its radius is estimated as $R_{\rm BLR}\p = 6.7\times 10^{17}\,$cm \citep{pft05}. Reference \cite{mea11} reported a tentative detection of a dusty torus (DT) with a luminosity $L_{\rm DT}\p = 7.0\E{45}\,$erg/s. Scaling relations (e.g., \citep[][]{Hea12}) then provide a radius of $R_{\rm DT}\p = 6.18\E{18}\,$cm$\,\sim 2\,$pc. The mass of the central black hole is estimated at $M_{\rm bh}\sim 8.5 \times 10^8\, M_{\odot}$~\citep{zcst14} giving an Eddington luminosity of $L_{\rm Edd}\p\sim 1.1\times 10^{47}\,$erg/s. \cta's accretion disk luminosity is $L_{\rm disk}\p = 3.8\times 10^{46}\,$erg/s$\,\sim 35\%\,L_{\rm Edd}$ \citep{zcst14}. All these quantities are given in the host galaxy frame.

\cta\ has attracted significant attention in recent years owing to a tremendous outburst in almost all energy bands \cite{bea17,zea17,rea17,sea18,gea18,kb18,pea18,zea19,cea19}. Over a period of four months, the flux rose and fell steadily (not counting short-term bursts on top of the trend) by up to two orders of magnitude. The short-term bursts revealed intra-night variability, potentially as short as a few minutes. The optical magnitude, typically at $R\sim 16\,$mag, reached a maximum of $R\sim 10.7$. {In turn, \cta\ became} visible by eye through small telescopes.

\begin{figure}[H]

\centering
\includegraphics[width=0.3\textwidth]{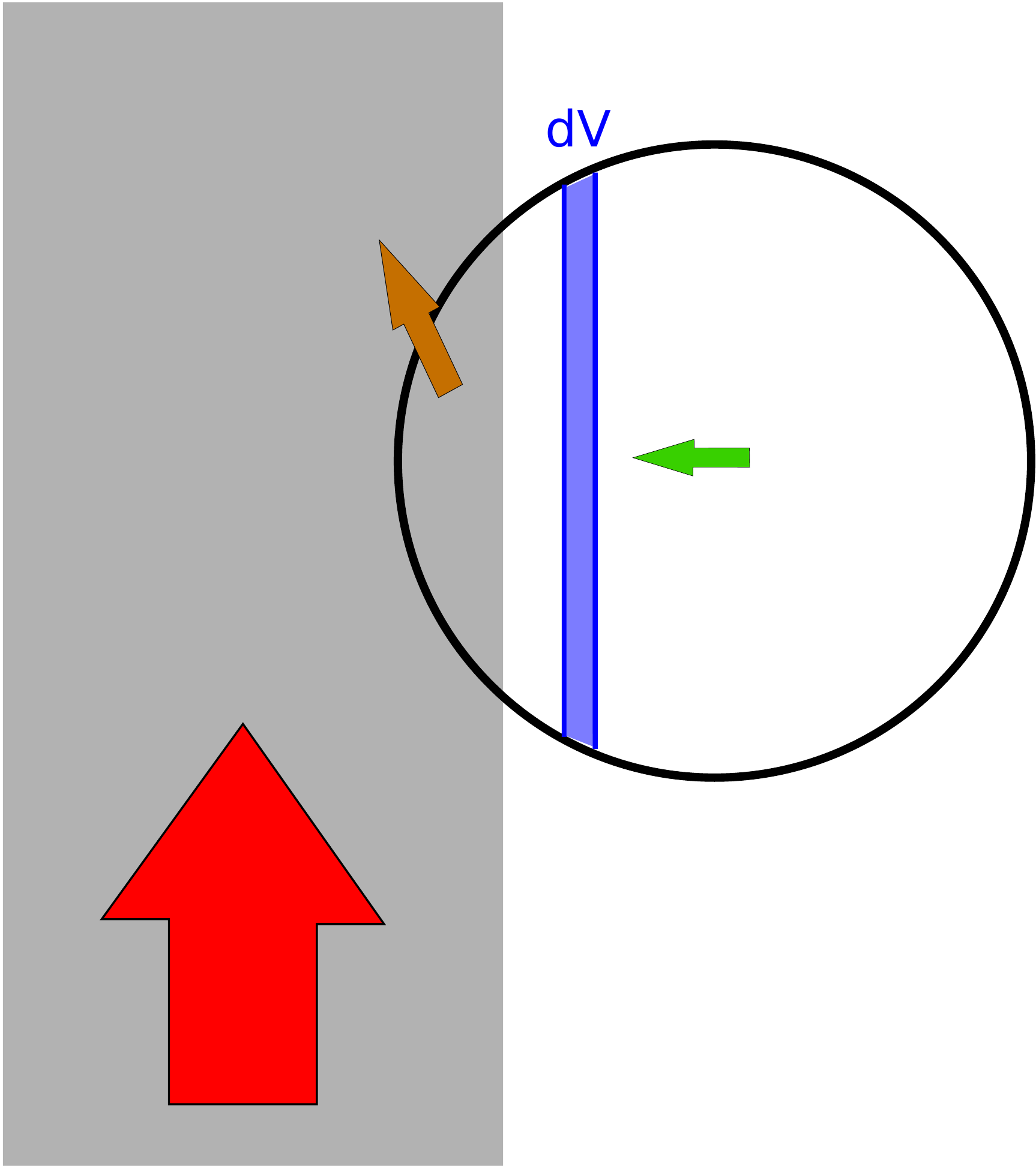}
\caption{Sketch of the model. The jet's ram pressure (red arrow) ablates the material that has entered the jet (brown arrow). The differential volume $\td{V}$ entering the jet in a differential time interval $\td{t}$ is given by the blue area. The green arrow marks the (horizontal) motion of the cloud with respect to the jet \cite{zea19}.}
\label{fig:sketch}
\end{figure}

Here, we summarize a model that might explain the four month outburst and the long-term trend around it. All details are given in \cite{zea17,zea19}. It is based on the idea of the ablation of material from an object penetrating the relativistic jet (e.g., \cite[][]{bab10,bpb12,dcea17}). The object may be a star or some other object, and we refer to it as a ``gas cloud''. The high ram pressure of the jet easily overcomes the gravitational pull of the cloud, and strips off any material that enters the jet. A sketch is shown in Figure~\ref{fig:sketch}.

%
%

\section{Results}
\begin{figure}[H]
\centering
\includegraphics[width=0.75\textwidth]{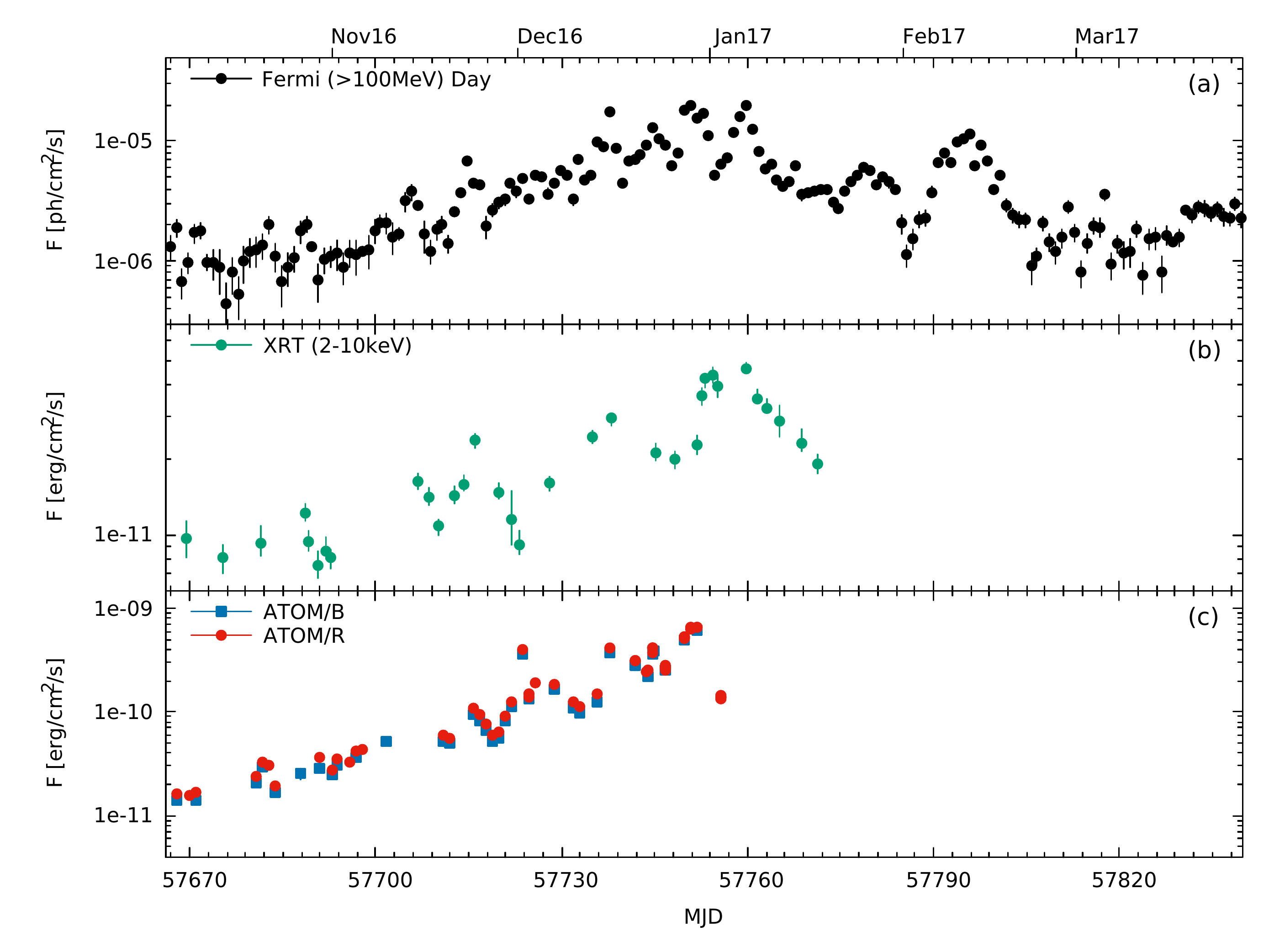}
\caption{{Multiwavelength light curve} in daily bins of \cta\ during the outburst in late 2016 and early 2017. Note the logarithmic scaling of the y-axis \cite{zea17}. 
}
\label{fig:mwl-lc}  

\end{figure}

The data analysis has been done by Zacharias et al. \cite{zea17,zea19}, and the reader is referred to these papers for details. We have analyzed data from the \fermi\ instrument, from the Neil Gehrels Swift Observatory and from ATOM.
The multiwavelength light curves are shown in Figure~\ref{fig:mwl-lc}. They indicate correlated behavior between the the optical, X-ray, and $\gamma$-ray fluxes. This strongly suggests that the emission is co-spatial. Unfortunately, the second half of the flare is not covered by any instrument except \fermi, as the sun blocks the view of \cta\ in the early months of a year. \footnote{Data obtained using optical observations from the northern hemisphere, indicate that the optical light curve followed a similar, symmetric trend as the $\gamma$-ray flux past January 2017 \cite{lea17}.} We have also analyzed the data from the three instruments obtained since mid-2008---the launch of {\it Fermi}---until 1 November 2018. The resulting light curve is shown in Figure~\ref{fig:longterm}.

The 2016--2017 outburst has been modeled using the model described briefly above. \mbox{The continuous} ablation of the gas cloud results in a time-symmetric injection of particles into the jet:
\begin{align}
 Q_{\rm inj}(t) \propto \logb{\frac{t_0^{\prime 2} + t_c^{\prime 2}}{t_0^{\prime 2}+(t_c\p-t\p)^2}} \label{eq:injection} .
\end{align}

Here, the time scales are related to the geometry of the cloud. $t_c$ is the time half of the cloud needs to penetrate the jet, and is thus related to the cloud's outer radius. $t_0$ is related to the cloud's scale height $r_0\propto \sqrt{T_{\rm c}/n_0}$ with the temperature $T_{\rm c}$ and the central density $n_0$ of the cloud.  {The jet parameters are fixed through modeling the pre-flare spectrum. Hence, the variability depends only on three free parameters, namely $t\p_c$, $t\p_0$, and the injection luminosity of the variability. $t\p_c$ is fixed by the flare duration, while the injection luminosity is related to the luminosity difference. $t\p_0$ governs the form of the light curve, and was varied until the light curve was nicely reproduced ({The consequences of varying the scale height will be the topic of a forthcoming publication.})---$t\p_0$ is, hence, the only true free parameter that is not directly inferred from the data. As the density $n_0$ is directly related to the variability injection luminosity, the temperature of the cloud can be derived.} Further details about the model and the radiative code used for modeling of the data are given in \cite{zea17,zea19}.

The injected particles get accelerated at a shock. This could be a shock resulting from the jet-cloud interaction or at a standing shock downstream in the jet. For the interpretation (see below) we have assumed that the acceleration and radiation zones in the jet are quasi-co-spatial with the interaction zone---but this is not a necessity. The accelerated particles, whether they are electrons or protons, interact with the ambient magnetic and photon fields to produce the observed radiation. A purely leptonic model is confined to within the extension of the BLR, as otherwise not enough photons are available to produce the inverse Compton signal. A hadronic model can, in principle, be placed at any distance from the black hole, as it is less dependent on external photon fields.

\begin{figure}[H]

\centering
\includegraphics[width=0.7\textwidth]{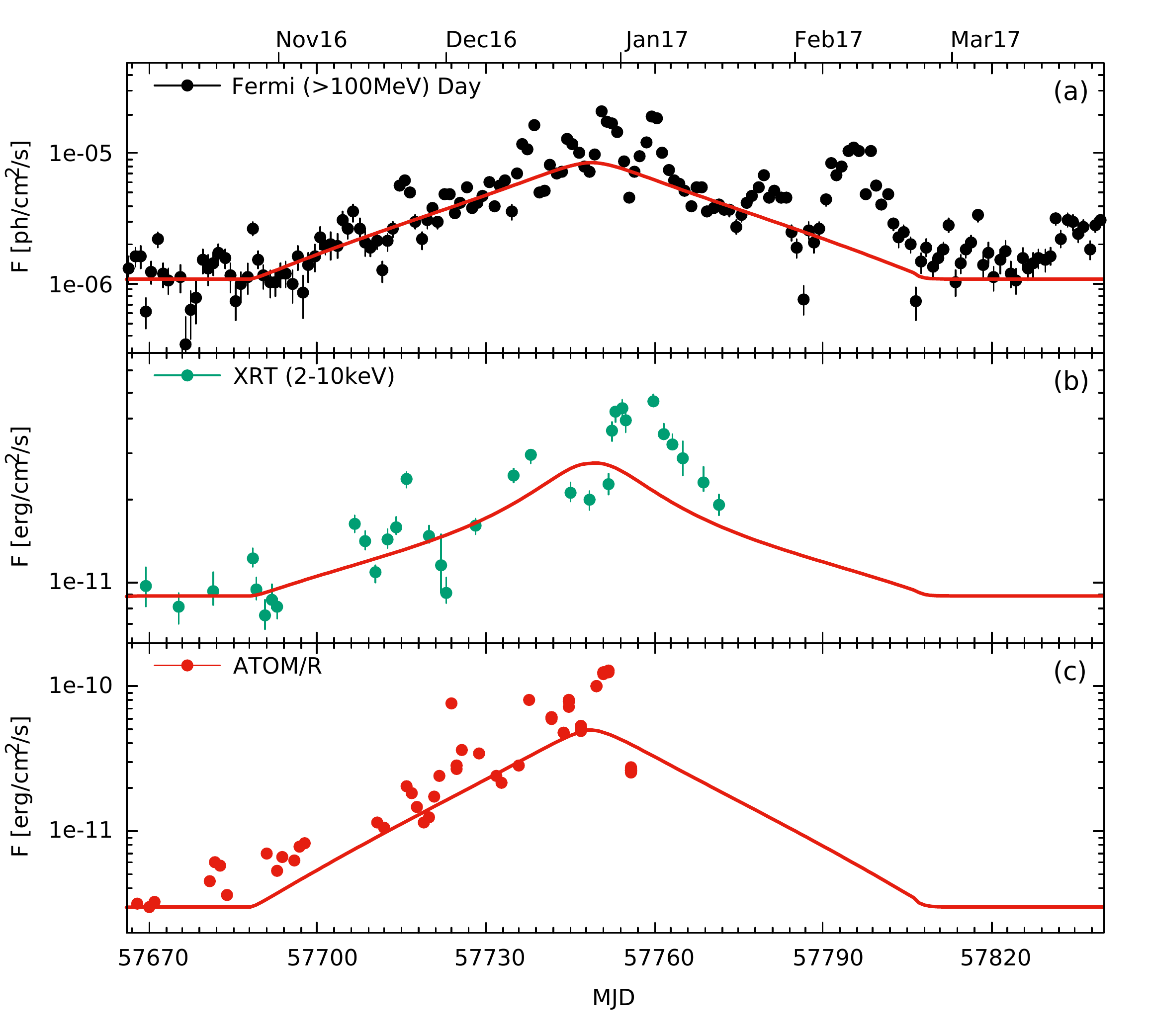}
\caption{{Model light curves} (red curves) using a leptonic model within the  broad-line region (BLR) for the $\gamma$-ray, X-ray and R-band \cite{zea17}. 
}
\label{fig:mod-lep}
\end{figure}

\begin{figure}[H]

\centering
\includegraphics[width=0.7\textwidth]{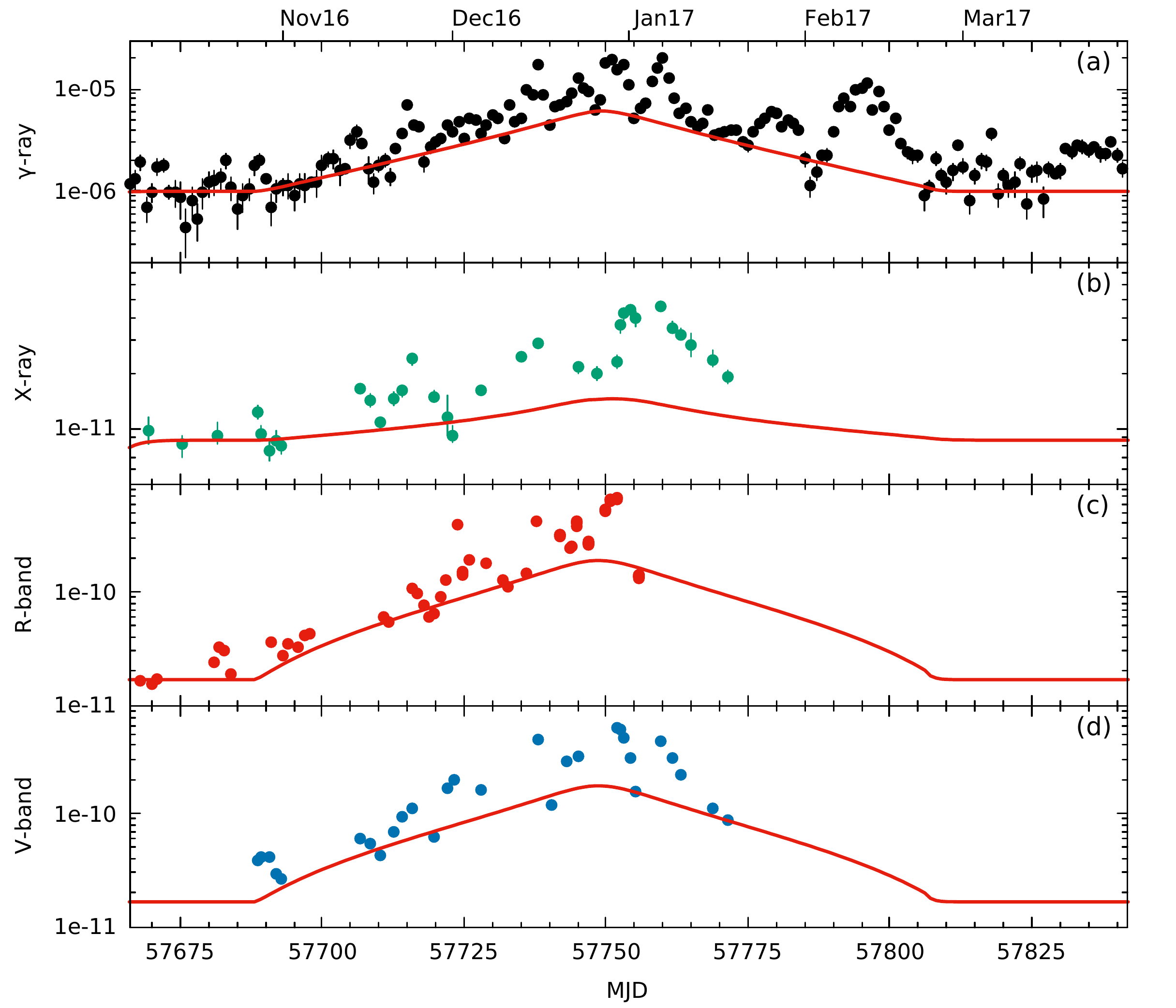}
\caption{{Model light curves} (red curves) using a hadronic model within the dusty torus (DT) for the $\gamma$-ray, X-ray, R- and V-band \cite{zea19}. 
}
\label{fig:mod-had}

\end{figure}

Figure~\ref{fig:mod-lep} shows the results of the leptonic model, while Figure~\ref{fig:mod-had} shows the same for a hadronic model placed within the DT but beyond the BLR. Both models fit the long-term behavior of the four month flare equally well. However, the two cases result in different parameters of the incoming cloud, which can be inferred from the parameters used for the modeling. The cloud parameters are given in Table~\ref{tab:cloud}.
\unskip

\begin{table}[H]
\centering
\caption{Inferred cloud parameters from the leptonic and hadronic modeling shown in Figures~\ref{fig:mod-lep} and \ref{fig:mod-had}.}
\begin{tabular}{lclcc}
\toprule
\multicolumn{3}{l}{\textbf{Model}} 				& \textbf{Leptonic}	& \textbf{Hadronic}   \\
\midrule
Distance from black hole	& $D$	& [cm]		& $6.5\E{17}$	& $3.1\E{18}$ \\
Cloud speed		& $v_{\rm c}$	& [cm/s]	& $5.1\E{8}$	& $1.9\E{8}$ \\
Cloud radius		& $R_{\rm c}$	& [cm]		& $1.3\E{15}$	& $4.9\E{14}$ \\
Cloud density		& $n_{\rm c}$	& [cm$^{-3}$]	& $2.5\E{8}$	& $1.1\E{7}$ \\
Cloud mass		& $M_{\rm c}$	& [g]		& $3.9\E{30}$	& $9.1\E{27}$ \\
Cloud temperature	& $T_{\rm c}$	& [K]		& $0.5$		& $2.7\E{-3}$\\
\bottomrule
\end{tabular}
\label{tab:cloud}
\end{table}
The speed is assumed to be the Keplerian speed around the black hole, and thus depends on the distance from the black hole. For the given distances, the gravitational potential is dominated by the black hole, and the assumption is well justified. As the cloud's radius depends on the cloud's speed and the duration of the flare, the radius shrinks with increasing distance. Within a hadronic model, less particles are required to produce the same radiative output, as we used a substantially higher magnetic field of $60\,$G as opposed to $3.7\,$G in the leptonic model. On the other hand, the density of the cloud increases with distance from the black hole, as in a given model the number of particles is roughly constant, while the radius shrinks. The mass $M_{\rm c}$ has been calculated assuming a pure hydrogen plasma in the cloud. Note that the cloud density $n_{\rm c}$ given in Table~\ref{tab:cloud} is the average density in the cloud and not the central density $n_0$ mentioned above. The temperature $T_{\rm c}$ is determined from the scale height $r_0$, which is a free parameter in our model, using the proportionality given above.

%
\section{Discussion}
While the modeling of the data is successful in both the leptonic and the hadronic case, \mbox{the inferred} cloud parameters raise questions in both cases. First of all, how could a cloud be so cold? And secondly, what kind of objects could fulfill these parameters?

The cloud temperature has been inferred from the cloud density, which in turn is inferred from the required particles additionally injected into the jet to cause the outburst. However, this line of inferences assumes that all particles of the cloud are injected into and accelerated in the jet. This~efficiency of 100\% is, however, highly unlikely. Observations of supernova remnants indicate that less than 10\% of the swept-up particles are accelerated in the shock front. This would directly increase the number of particles, and thus the temperature 10-fold. Furthermore, the interaction of the cloud with the jet has been described as a smooth process. This, as well, is highly unlikely. The energy density of the jet is so much higher than the cloud's energy density that the jet looks like a giant wall to the cloud. \footnote{As an analogy, imagine to try to drop a droplet of water into the water jet from a garden hose. Most of the droplet is scattered elsewhere.} Hence, the impact of the cloud would eject most cloud particles elsewhere, while only a few particles might actually enter the jet. Statistically, this would still follow the injection form given above. While both these effects imply that the original density and temperature of the cloud are much higher than what we have inferred from the modeling, this is difficult to quantify. Hence, for the continuing discussion about the nature of the cloud, we stick to the inferred parameters.

Cloud-like structures are ubiquitous in galaxies. We concentrate on the following possibilities as these are the densest structures and could be abundant in AGN  hosts: Clouds of the BLR, molecular clouds forming stars, and stellar atmospheres. The diffuse gas that could be present even in an elliptical galaxy such as the host of \cta, is too thin to produce a significant signal.

BLR clouds are usually given with a radius of $\sim 10^{13}\,$cm and densities of $10^{9..11}\,$cm$^{-3}$ 
\cite{dea99,p06} . While the densities might be possible with our clouds provided the given densities in Table~\ref{tab:cloud} are lower limits, the radius of a cloud within the BLR (see the leptonic model) is too large compared to BLR cloud radii. This makes BLR clouds unlikely candidates  {for this flare. Given the size and typical velocities of BLR clouds, flares induced by BLR clouds should last for a day or less. This could explain the more common short-term flares \cite{abr10}.} On the other hand, observations suggest that the BLR is not isotropically distributed (e.g., \cite[][]{gea13}), so interactions of BLR clouds with the jet  {may not happen very~often.}

Star-forming cloud cores, embedded in giant molecular clouds, can reach sizes of  0.05--1 pc and densities of $10^{7..9}\,$cm$^{-3}$ (e.g., \citep[][]{co17}, chp. 12). This fits well with the densities, however our clouds are too small. On the other hand, the density is an average over the core, which hosts several forming stars. Therefore, the actual star formation sites are smaller, and might therefore work as a seed for this flare, as well.  {As elliptical galaxies, as the hosts of blazars, are typically devoid of significant gas reservoirs, there might not be many star forming regions in a blazar host galaxy. This could explain the singularity of the discussed event. If the host of \cta\ contains gas and/or star forming regions requires dedicated observations.}

Lastly, we consider the astrospheres of red giant stars. These are post main-sequence stars with highly inflated atmospheres. The atmospheres can reach radii of $10^{13}\,$cm. This is one or two orders of magnitude too small for the outburst. On the other hand, as mentioned already these radii depend strongly on the velocity of the cloud. If the object would be moving with $\sim$100 km/s, the radius would fit nicely with our model. The given velocity is a typical one for stars in large elliptical galaxies. Hence, the interaction of a red giant star with the jet at large distances from the black hole is a plausible scenario.  {An elliptical galaxy hosts mostly red stars, and a significant number of them may be in the red-giant phase at any given time. How many of these are also on an orbit that crosses the relativistic jet, is difficult to say precisely. However, a simple estimate may be as follows. A typical elliptical galaxy may have a diameter radius of $\sim$100 kpc and host about $10^{11}$ stars. In order for the ablation model to work, the jet's ram pressure needs to be strong enough, which it may be within the first $\sim$100~pc from the black hole. Less than $\sim$10 stars would be located within this volume of the jet. Red~giants constitute less than $1\%$ of all stars in a galaxy as that evolutionary phase is relatively short. Hence, the interaction of red giants with the inner jet may be rare events. }

\begin{figure}[H]
\centering
\includegraphics[width=0.9\textwidth]{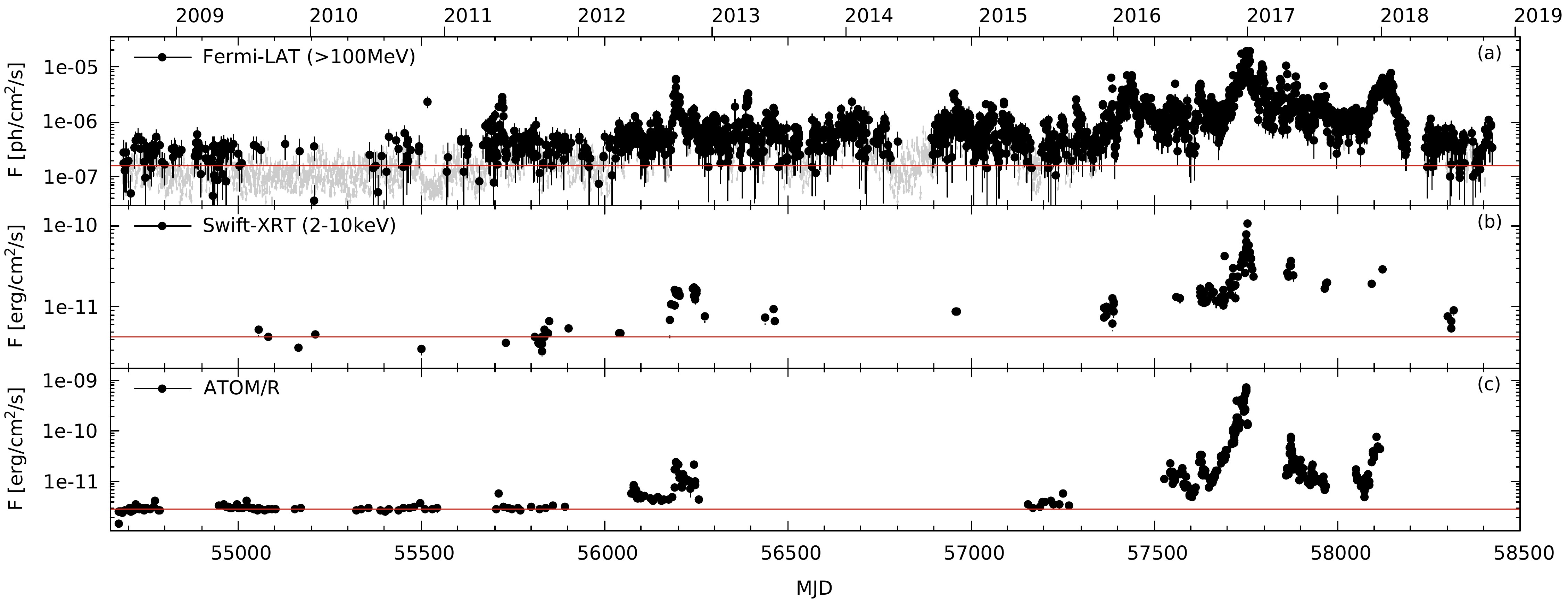}
\caption{{Long-term multiwavelength} light curve in daily bins of \cta\ since mid-2008. Note the logarithmic scaling of the y-axis. The gray arrows mark upper limits and the red lines the pre-2012 averages \cite{zea19}. 
}
\label{fig:longterm}
\end{figure}

So far, the true nature of the cloud cannot be convincingly determined. However, it is possible that the source behavior on even longer scales might give a hint. We have therefore analyzed all data in $\gamma$-rays from \fermi, X-rays from Swift-XRT and optical R-band from ATOM since mid-2008. The~data is shown in Figure~\ref{fig:longterm}. The $\gamma$-ray data indicates that the source evolution can be divided into 4 broad time frames. The first one lasted until early-2012 characterized by a low flux state, where the source has rarely been detected in daily bins. From 2012 till the beginning of 2016 the average flux has significantly increased and the source became detectable for most daily bins. From 2016 till the end of 2018 the source average has increased further. The beginning and end of this phase is marked by strong flares. The four months outburst discussed in detail above is in the middle of this period. Since~early 2018 the source has seemingly returned to pre-2016 levels.	The X-ray and R-band light curves support these statements---however, their sparse observation cadences do not allow one to draw firm~conclusions.

The symmetry in the two year event from 2016 to 2018 is striking and may point towards a common origin of the entire event encompassing the four months outburst. Without a detailed modeling it is difficult to draw conclusions, which of the discussed scenarios for the cloud origin is more plausible. The symmetry points towards an individual object, such as a red giant star, while the length of the event may support an extended object, such as a molecular cloud with embedded star forming regions. Nonetheless, this is fascinating enough to keep wondering about the nature of this truly singular event.

{In fact, we are not aware of another similar event in the last couple of years. While \fermi\ has observed many bright flares, those usually were on much shorter time scales. A couple of long-lasting high states have also been detected, but none of them has produced comparable fluxes. It might be that a specific configuration of jet and cloud parameters was necessary to produce the outburst in \cta, and that slightly different configurations lead to different light curve profiles and fluxes. As~the nature of the cloud remains unknown, it is difficult to draw conclusions on this particular event, and the discussion of light curves expected from different clouds is beyond the scope of this paper. }

%
\section{Conclusions}
We have summarized the modeling of an extended, four months-long outburst in the FSRQ \cta. The model employs the ablation of a gas cloud by the relativistic jet. As the jet's ram pressure easily overcomes the cloud's gravitational pull, this process is gradual and begins immediately after the first contact of the cloud with the jet. As the ablated volume in a given time interval along with the cloud's density changes with time (aka, as the cloud further penetrates the jet), the injection of cloud particles into the jet is a time-dependent, but symmetric event. As shown by \mbox{Zacharias et al. \cite{zea17,zea19}}, this~injection model can easily explain the four months outburst with its symmetric light curve. Both~leptonic and hadronic models can account for the event.

This is important, as the hadronic model provides greater freedom with respect to the distance of the event from the black hole. The leptonic model is confined to take place within the boundaries of the BLR. However, cloud parameters inferred from the modeling do not fit well with BLR cloud parameters. The hadronic model can be placed at any distance from the black hole, as it is not as dependent on the external photon fields as the leptonic model. Parameters inferred from the hadronic modeling might fit with both the dense star-forming cores of molecular clouds or the atmosphere of a red giant star. If these possibilities could also explain the symmetric two year event that underlines the four months outburst in \cta\ is an intriguing, yet open question.

Nonetheless, the fascinating nature of this event underlines the need for highly detailed monitoring of a large number of sources. Only then, such events and also their evolution around the main event can be properly analyzed and interpreted.


\vspace{6pt}



\authorcontributions{Conceptualization, M.Z.; Data curation, F.J., J.-P.L. and A.W.; Formal analysis, F.J., J.-P.L. and A.W.; Funding acquisition, S.W.; Investigation, M.Z.; Software, M.B., F.J., J.-P.L. and A.W.; Validation, S.W.; Writing---original draft, M.Z.; Writing---review \& editing, M.B. and S.W.
}

\funding{M.Z. acknowledges funding by the German Ministry for Education and Research (BMBF) through grant 05A17PC3.
The work of M.Z. and M.B. is supported through the South African Research Chair Initiative (SARChI) of the South African Department of Science and Technology (DST) and National Research Foundation. \footnote{Any opinion, finding and conclusion or recommendation expressed in this material is that of the authors, and the NRF does not accept any liability in this regard.}.
F.J. and S.J.W. acknowledge support by the German Ministry for Education and Research (BMBF) through Verbundforschung Astroteilchenphysik grant 05A11VH2.
A.~W. is supported by the Foundation for Polish Science (FNP).
}

\acknowledgments{J.-P.~L. gratefully acknowledges CC-IN2P3 (\href{https://cc.in2p3.fr}{cc.in2p3.fr}) for providing a significant amount of the computing resources and services needed for this work.
}

\conflictsofinterest{The authors declare no conflict of interest.
}

%

\appendixtitles{no} 
\appendixsections{multiple} 
%


\reftitle{References}





\end{document}